\documentclass[aps,pra,twocolumn,superscriptaddress,showpacs,amsmath,amssymb]{revtex4-2}
\usepackage[breaklinks=true,colorlinks,citecolor=blue,linkcolor=blue,urlcolor=blue]{hyperref}
\usepackage{graphicx}
\usepackage{color}
\usepackage{braket}
\usepackage{siunitx}
\usepackage{bm}
\usepackage{makecell}
\usepackage{caption}
\usepackage{subfig}
\usepackage[normalem]{ulem}
\usepackage[usenames,dvipsnames]{xcolor}
\usepackage{chemformula}

\begin{document}

\title{Post-processing noisy quantum computations utilizing N-representability constraints}

\author{Tomislav Piskor}
\email{tomislav.piskor@eviden.com}
\affiliation{HQS Quantum Simulations GmbH, Rintheimer Strasse 23, 76131 Karlsruhe, Germany}
\affiliation{Theoretical Physics, Saarland University, 66123 Saarbr\"ucken, Germany}
\affiliation{science + computing AG / Eviden, Hagellocher Weg 73, 72070 Tübingen, Germany}

\author{Florian G. Eich}
\affiliation{HQS Quantum Simulations GmbH, Rintheimer Strasse 23, 76131 Karlsruhe, Germany}

\author{Michael Marthaler}
\affiliation{HQS Quantum Simulations GmbH, Rintheimer Strasse 23, 76131 Karlsruhe, Germany}

\author{Frank K. Wilhelm}
\affiliation{Theoretical Physics, Saarland University, 66123 Saarbr\"ucken, Germany}
\affiliation{Institute for Quantum Computing Analytics (PGI-12), Forschungszentrum J\"ulich, 52045 J\"ulich, Germany}

\author{Jan-Michael Reiner}
\affiliation{HQS Quantum Simulations GmbH, Rintheimer Strasse 23, 76131 Karlsruhe, Germany}

\begin{abstract}
We propose and analyze a method for improving quantum chemical energy calculations on a quantum computer impaired by decoherence and shot noise.
The error mitigation approach relies on the fact that the one- and two-particle reduced density matrices (1- and 2-RDM) of a chemical system need to obey so-called N-representability constraints.
We post-process the result of an RDM measurement by projecting it into the subspace where certain N-representability conditions are fulfilled.
Furthermore, we utilize that such constraints also hold in the hole and particle-hole sector and perform projections in these sectors as well.
We expand earlier work by conducting a careful analysis of the method's performance in the context of quantum computing.
Specifically, we consider typical decoherence channels (dephasing, damping, and depolarizing noise) as well as shot noise due to a finite number of projective measurements.
We provide analytical considerations and examine numerically three example systems, \ch{H2}, \ch{LiH}, and \ch{BeH2}.
From these investigations, we derive our own practical yet effective method to best employ the various projection options.
Our results show the approach to significantly lower energy errors and measurement variances of (simulated) quantum computations.
\end{abstract}

\maketitle

\section{Introduction}
\label{sct:intro}
Recently, the steadily advancing development of quantum computers has demonstrated more and more the potential of this emerging technology.
Even though we are still in the era of noisy intermediate scale quantum (NISQ) devices~\cite{preskill_quantum_2018}, a number of experiments showed highly promising results regarding possible advantages of quantum computing over conventional computing~\cite{arute_quantum_2019, zhong_quantum_2020, wu_strong_2021, zhu_quantum_2021, madsen_quantum_2022},
with one of the most promising field of future application being in the realm of quantum simulations, e.g., of materials or chemical systems~\cite{abrams_simulation_1997, abrams_quantum_1999}, with impressive recent demonstrations~\cite{mazurenko_cold-atom_2017, kandala_hardware-efficient_2017, kandala_error_2019, arute_observation_2020, rost_demonstrating_2021, tazhigulov_simulating_2022, xu_digital_2022, guo_experimental_2023}.

Yet, at the current stage, quantum resources are fairly limited.
Despite tremendous achievements concerning quantum error correction~\cite{erhard_entangling_2021, acharya_suppressing_2023}, current qubit systems are too small for the application of quantum error correction on a sufficient scale~
\cite{fowler_surface_2012, lekitsch_blueprint_2017, bermudez_assessing_2017}, leading to erroneous results due to decoherence.
Furthermore, error-prone measurements are another source of inaccuracy for algorithms discussed for NISQ devices;
particularly for variational algorithms~\cite{farhi_quantum_2014, peruzzo_variational_2014, wecker_progress_2015, mcclean_theory_2016} where a large number of measurements needs to be taken during the challenging optimization procedure of the algorithm's variational parameters~\cite{mcclean_barren_2018, kubler_adaptive_2020, sweke_stochastic_2020, meyer_variational_2021, piskor_using_2022}.

In this work, we present and analyze a method for improving the accuracy of a quantum chemical simulation on a noisy quantum computer by means of post-processing.
We are interested in calculations where the energy of a system in a given state is evaluated via measuring the expectation values $\braket{c_i^\dagger c_j}$ and $\braket{c_i^\dagger c_j^\dagger c_l c_k}$, i.e., the elements of the one- and two-particle reduced density matrix (1- and 2-RDM), where $c^{(\dagger)}_i$ are fermionic an\-ni\-hi\-la\-tion (creation) operators of the system in second quantization.
On a NISQ computer, we will obtain flawed values for these elements.
However, we know that the 1- and 2-RDM need to satisfy certain conditions and, hence, could try to mitigate the error by post-processing the result accordingly, to conform with these constraints.

Specifically, we consider the so-called N-re\-pre\-sen\-ta\-bi\-li\-ty constraints that the RDMs need to obey~\cite{mazziotti_structure_2012, mazziotti_significant_2012}.
They originate from the N-representability problem posed when trying to guarantee a 2-RDM remains derivable from a valid density matrix of $N$ fermions, while modifying the 2-RDM in a variational approach to minimize the energy~\cite{mayer_electron_1955, coleman_structure_1963}.
Using N-representability constraints to improve on a quantum chemistry calculation (on conventional computers) in a post-processing manner was previously suggested~\cite{lanssens_method_2018}, and also considered in a quantum computing context~\cite{rubin_application_2018}.

In these two references~\cite{lanssens_method_2018, rubin_application_2018}, the concept is to pick a subset of the N-representability constraints and project an erroneous RDM from a calculation to the (in the Frobenius norm) closest matrix that fulfills the selected constraints to obtain an improved RDM.
Furthermore, this projection is not solely performed for the two-particle RDM (with elements $\braket{c_i^\dagger c_j^\dagger c_l c_k}$), but also the two-hole RDM ($\braket{c_i c_j c_l^\dagger c_k^\dagger}$), and particle-hole RDM ($\braket{c_i^\dagger c_j c_l^\dagger c_k}$).

We expand on this work by providing a more thorough analysis for the application of this concept in energy calculations using a quantum computer, enhancing the understanding of the procedure and its usefulness, and enabling us to formulate new practical modifications of the method.
We add an investigation how different sources of noise in a quantum computation, individually and in combination, affect the measured RDMs.
This is done through analytical consideration and, in particular, numerical simulation.
Specifically, we study the several combinations of noise and the three options of projecting in the respective particle, hole, or particle-hole sector, to observe where which projection enhances the result to which extent depending on the scenario.
As test systems, we consider three molecules, \ch{H2}, \ch{LiH} and \ch{BeH2}, and we consider three quantum noise channels, dephasing, damping, and depolarization, and furthermore shot noise, stemming from performing a limited number of measurements to evaluate the expectation values for the RDM elements.
Based on our results and new data, we propose a practical approach how to utilize having multiple sectors as options to perform the projection in.

This paper is structured as follows:
In Sec.~\ref{sct:theory} we give a short overview on the basic principles of N-representability as well as present notations that are valid throughout this work.
Sec.~\ref{sct:noise} explains how we are simulating noise via the superoperator formalism and also shows the operators for the three investigated decoherence types: dephasing, damping and depolarization. Furthermore, we give a brief description how we were simulating shot noise in combination with decoherence, and comment on how each type of noise leads to states violating N-representability.
A thorough description of our simulation and post-processing procedure, followed by the definitions of the considered metrics is given in Sec.~\ref{sct:method}.
Finally, Sec.~\ref{sct:results} gives a presentation of the results of our numerical analysis, before we conclude in Sec.~\ref{sct:conclusion}.

\section{Quantum chemistry and N-representability}
\label{sct:theory}
The aim of this work is, by the means of post-processing, to improve the accuracy of ground state energy calculations for molecular systems performed on a noisy intermediate scale quantum (NISQ) computer. We consider systems described by a spin-separated molecular Hamiltonian
\begin{equation}
	\mathcal{H} = \text{const.} + \sum_{ij}h_{ij} c_i^\dagger c_j + \frac{1}{2} \sum_{ijkl} V_{ijkl} c_i^\dagger c_j^\dagger c_l c_k,
	\label{eq:ham}
\end{equation}
where $\text{const.}$ collects all non-electron effects such as the interaction between the nuclei, $c^{(\dagger)}$ denotes the annihilation (creation) operators of the spin orbitals, and with the one- and two-electron tensors $h_{ij}$ and $V_{ijkl}$:
\begin{subequations}
	\begin{align}
		\label{eq:hij}
		h_{ij} = \int \mathrm{d}r \, \phi_i^*(r)\left(-\frac{\nabla^2}{2m} + \sum_I \frac{Z_I}{\left\lvert r-R_I\right\rvert}\right)\phi_j(r), \\
		\label{eq:vijkl}
		V_{ijkl} = \int \mathrm{d}r \mathrm{d}r^\prime \, \frac{\phi_i^*(r)\phi_j^*(r^\prime)\phi_k(r^\prime)\phi_l(r)}{\left\lvert r-r^\prime\right\rvert}.
	\end{align}
\end{subequations}
Here, $h_{ij}$ contains all one-electron effects such as the kinetic energy and the Coulomb interaction between the electron and the nuclei, where $\phi(r)$ denotes the spatial basis function. The two-electron integral $V_{ijkl}$ describes the Coulomb interaction between an electron located at position $r$ and an electron located at $r^\prime$.

The energy of a state $\ket\psi$ with respect to this system is given by the expectation value of the Hamiltonian~\eqref{eq:ham},
\begin{align}
	\begin{split}
		E = \langle \mathcal{H} \rangle &= \text{const.} + \sum_{ij} h_{ij} \braket{c_i^\dagger c_j} + \frac{1}{2} \sum_{ijkl} V_{ijkl} \braket{c_i^\dagger c_j^\dagger c_l c_k}\\
		&= \mathrm{const.} + \sum_{ij} h_{ij} {^1D^i_j} + \frac{1}{2} \sum_{ijkl} V_{ijkl} {^2D^{ij}_{kl}},
	\end{split}
	\label{eq:energy}
\end{align}
where we introduced $\langle \cdot \rangle = \bra\psi \cdot \ket\psi$ as shorthand notation, and the one-particle and two-particle reduced density matrices (1-RDM and 2-RDM):
\begin{subequations}
	\begin{align}
		\label{eq:1rdm}
		^1D^i_j & = \braket{c_i^\dagger c_j} = \bra{\psi} c_i^\dagger c_j \ket{\psi}, \\
		\label{eq:2rdm}
		^2D^{ij}_{kl} & = \braket{c_i^\dagger c_j^\dagger c_l c_k} = \bra{\psi} c_i^\dagger c_j^\dagger c_l c_k \ket{\psi}.
	\end{align}
\end{subequations}

When calculating the energy of a state on a quantum computer, one would map the fermionic operators onto qubits, e.g., using the Jordan-Wigner transformation, and measure the elements of the RDM.
However, with a NISQ device in particular and finite computational resources one will obtain an erroneous result due to decoherence and shot noise (more on our considered noise types and their respective descriptions in Sec.~\ref{sct:noise}).
In this work we analyze how we can reduce the error by post-processing the result where we utilize knowledge about certain constraints that the 1- and 2-RDM need to fulfill.
Specifically, we utilize the fact that an RDM needs to obey the so-called N-representability conditions if it is derived from a proper state of $N$ fermions~\cite{mazziotti_structure_2012}.
There is a variety of these conditions, especially when specifying particle and orbital numbers~\cite{smith_nrepresentability_1965, smith_n-representability_1966, mazziotti_significant_2012}; but in this work we focus on just a few constraints which generally hold for any fermionic systems with a well defined particle number (note that this is the case in chemical electronic structure problems):
\begin{enumerate}
	\item \textbf{Hermiticity} -- It is easy to see from Eqs.~\eqref{eq:1rdm} and~\eqref{eq:2rdm} that the 1- and 2-RDM are Hermitian, meaning that:
	\begin{subequations}
		\begin{align}
			^1D^i_j &= (^1D^j_i)^*, \label{eq:hermiticity}\\
			^2D^{ij}_{kl} &= (^2D^{kl}_{ij})^*.
			\label{eq:hermiticity2}
		\end{align}
	\end{subequations}
	\item \textbf{Antisymmetry} -- Making use of fermionic anticommutation relations, we can rewrite the elements of the 2-RDM:
		\begin{equation}
			{^2D}^{ij}_{kl} = -{^2D}^{ji}_{kl} = -{^2D}^{ij}_{lk} = {^2D}^{ji}_{lk}.
			\label{eq:antisymmetry}
		\end{equation}
	\item \textbf{Positive semidefiniteness} -- The 1- and 2-RDM has to be positive semidefinite, meaning that all eigenvalues of the matrices have to be non-negative.	
	\item \textbf{Trace integrity} -- From their definition one can derive that for a state with a well defined particle number $N$ the traces of the 1- and 2-RDM are given by:
	\begin{subequations}
		\begin{align}
			\sum_i {^1}D^i_i &= N, \label{eq:trace1}\\
			\sum_{ij} {^2}D^{ij}_{ij} &= N(N-1).
			\label{eq:trace2}
		\end{align}
	\end{subequations}
	\item \textbf{Contractibility} -- Related to the trace relation, for an $N$ particle state, one can find the 1-RDM elements by contraction of the 2-RDM:
	\begin{equation}
		^1D^i_j = \frac{1}{N-1}\sum_k {^2D}^{ik}_{jk},
		\label{eq:contraction}
	\end{equation}
\end{enumerate}

One should mention -- because we will exploit this later -- that these conditions not only hold for the one- and two-particle RDMs; similarly they also hold for the 1- and 2-hole, as well as the particle-hole RDMs:
\begin{subequations}
	\begin{align}
		^1Q^i_j &= \bra{\psi} c_i c_j^\dagger \ket{\psi}, \\
		^2Q^{ij}_{kl} &= \bra{\psi} c_i c_j c_l^\dagger c_k^\dagger \ket{\psi}, \\
		^2G^{ij}_{kl} &= \bra{\psi} c_i^\dagger c_j c_l^\dagger c_k \ket{\psi}.
		\label{eq:definitions}
	\end{align}
\end{subequations}
One can obtain these from the one- and two-particle RDMs using the following identities:
\begin{subequations}
	\begin{align}
		^1Q^j_i &= \delta_{ij} - ^1D^i_j,
		\label{eq:trans1Q1D} \\
		\begin{split}
			^2Q^{lk}_{ji} &= {^2D}^{ij}_{kl} - \delta_{jl} {^1D^i_k} + \delta_{il} {^1D}^j_k \\ &\quad + \delta_{jk} {^1D}^i_l - \delta_{ik} {^1D^j_l} + \delta_{jl} \delta_{ik} - \delta_{il} \delta_{jk}, 
		\end{split} \label{eq:trans2Q2D} \\
		^2G^{il}_{kj} &= \delta_{jl} {^1D}^i_k - {^2D}^{ij}_{kl}.
		\label{eq:trans2G2D}
	\end{align}
\end{subequations}

N-representability constraints were long utilized to improve quantum chemical calculations.
In fact, they stem from the N-representability problem posed when trying to guarantee a 2-RDM can be represented by, i.e., derived from, a proper state of $N$ fermions, while modifying the 2-RDM in a variational approach to minimize the energy~\cite{mayer_electron_1955, coleman_structure_1963}.
It was also proposed to look not just at the particle, but also the hole and particle-hole sectors to improve numerical methods~\cite{lanssens_method_2018}.
Furthermore, the constraints were applied in the context of quantum computing to improve measurement results of RDM elements~\cite{rubin_application_2018}, by projecting measured RDMs into the subspace where specific N-representability constraints were fulfilled.

We expand on this work, focusing on a quantum computing application, providing a new practical method that exploits N-representability conditions in the particle, hole, and particle-hole sector.
We derive our approach from performing a thorough analysis how individual noise types, and combinations of them, affect the performance of our method.
To this end, in the next section we continue to explain which kinds of noise we consider, how we describe them, and how the individual N-representability constraints above are affected by them.

\section{Noise}
\label{sct:noise}
In this work we consider three different types of decoherence noise, namely damping, depolarization and dephasing.
To include stochastic effects, we will also investigate shot noise.
In this section we will discuss the superoperator formalism which we utilize, as well as the single noise types and present how the individual noise types influence the N-representability conditions listed above.

\subsection{Decoherence}
\label{susct:decoherence}

To describe the effect of decoherence noise we make use of the superoperator formalism after vectorizing the density matrix~\cite{horn_johnson_1990,machnes_2014}. For each noise type there exists a corresponding superoperator, which will be defined in the following. Assuming a general one qubit density matrix $\rho$, we can transform it to a vectorized form
\begin{equation}
	\label{eq:vecdm}
	\rho = 
	\begin{pmatrix}
	\rho_{00} & \rho_{01} \\
	\rho_{10} & \rho_{11}
	\end{pmatrix}
	\rightarrow \vec{\rho} = \begin{pmatrix} \rho_{00} \\ \rho_{10}  \\ \rho_{01} \\ \rho_{11} \end{pmatrix}.
\end{equation}
As a next step, we apply the one-qubit superoperator $\mathcal L$ to the vectorized density matrix,
\begin{align}
	\label{eq:dephvecdm}
	\mathcal{L}\vec{\rho} = 
	\begin{pmatrix}
	1 & 0 & 0 & 0 \\
	0 & \mathrm{e}^{-2 \Gamma t} & 0 & 0 \\
	0 & 0 & \mathrm{e}^{-2 \Gamma t} & 0 \\
	0 & 0 & 0 & 1 \\
	\end{pmatrix}
	\begin{pmatrix} \rho_{00} \\ \rho_{10}  \\ \rho_{01} \\ \rho_{11} \end{pmatrix} = 
	\begin{pmatrix} \rho_{00} \\ \mathrm{e}^{-2 \Gamma t} \rho_{10} \\ \mathrm{e}^{-2 \Gamma t} \rho_{01} \\ \rho_{11} \end{pmatrix}\!,
\end{align}
where here we used the dephasing superoperator as an example, with $\Gamma$ denoting the dephasing rate and $t$ the time the noise acted on the system.
Transforming Eq.~\eqref{eq:dephvecdm} back to matrix form yields
\begin{equation}
	\label{eq:dephdm}
	\begin{pmatrix}
	\rho_{00} & \mathrm{e}^{-2 \Gamma t} \rho_{01} \\
	\mathrm{e}^{-2 \Gamma t} \rho_{10} & \rho_{11}
	\end{pmatrix}.
\end{equation}
Note, that going from pure states to mixed states, relying on density matrices Eqs.~\eqref{eq:1rdm} and~\eqref{eq:2rdm} would read:
\begin{align}
	^1D^i_j & = \braket{c_i^\dagger c_j} = \mathrm{Tr} (\rho c_i^\dagger c_j ), \\
	^2D^{ij}_{kl} & = \braket{c_i^\dagger c_j^\dagger c_l c_k} = \mathrm{Tr} (\rho c_i^\dagger c_j^\dagger c_l c_k ).
\end{align}
Here, $\mathrm{Tr}(\cdot)$ denotes the trace of the operator.

The superoperator formalism can be extended to multiple qubits.
Here, one vectorizes the multi-qubit density matrix and expands the superoperator such that it acts on the subspace of the respective qubit.
Throughout this work, we assume that each noise type affects all qubits equally.
Hence, we sequentially apply the superoperators acting on each individual qubit, with equal noise rates for all qubits (see Appx.~\ref{appx:multi-qubit} for a two-qubit example).

\emph{Dephasing} was just used as an example with the superoperator given in Eq.~\eqref{eq:dephvecdm}. Dephasing can be understood as random phase errors, i.e., Pauli $Z$ applications on qubits with a certain rate. Averaging over many random instances yields a density matrix equivalent to using the superoperator formalism.

The effect of \emph{damping noise} can be seen as the relaxation of a qubit with a certain probability, i.e., the qubit -- initially being in the excited state $\ket{1}$ -- decays after a certain amount of time to the ground state $\ket{0}$. The superoperator for the damping channel is given as:
\begin{equation}
	\mathcal{L} = \begin{pmatrix}
	1 & 0 & 0 & 1 - \mathrm{e}^{-\Gamma t} \\
	0 & \mathrm{e}^{-\frac{\Gamma}{2} t} & 0 & 0 \\
	0 & 0 & \mathrm{e}^{-\frac{\Gamma}{2} t} & 0 \\
    0 & 0 & 0 & \mathrm{e}^{-\Gamma t} \\
	\end{pmatrix}.
	\label{eq:SODamping}
\end{equation}
Here, $\Gamma$ denotes the damping rate.

\emph{Depolarization} can be seen as bit and phase flip errors acting on the qubits. The superoperator representation of this noise gate is defined as:
\begin{equation}
	\mathcal{L} = \begin{pmatrix}
	\frac{1}{2} (1 + \mathrm{e}^{-\Gamma t}) & 0 & 0 & \frac{1}{2} (1 - \mathrm{e}^{-\Gamma t}) \\
	0 & \mathrm{e}^{-\Gamma t} & 0 & 0 \\
	0 & 0 & \mathrm{e}^{-\Gamma t} & 0 \\
	\frac{1}{2} (1 - \mathrm{e}^{-\Gamma t}) & 0 & 0 & \frac{1}{2} (1 + \mathrm{e}^{-\Gamma t}) \\
	\end{pmatrix},
	\label{eq:SODepolarization}
\end{equation}
with the depolarizing rate $\Gamma$.

Note, that in our numerical simulations below, we set the evolution time to $t = 1$ and scale the rates $\Gamma$ accordingly in dimensionless units.

\subsection{Shot noise}
\label{subsct:shot}

Performing computations on a real quantum device requires multiple projective measurements of qubits in the computational basis in order to extract operator expectation values.
In this work we faithfully simulated this measurement process to obtain the expectation values influenced by shot noise using our software package \texttt{qoqo}~\cite{noauthor_github_nodate}.

The software does this by grouping the operators (Pauli products) to be measured into sets that can be measured simultaneously.
For each set, the quantum circuit is extended by the respective single-qubit rotations such that the Pauli products can be measured in the computational basis.
Each extended circuit is then simulated and the resulting final state vectors are obtained.
From there, for each measurement shot a bit string is drawn from a probability distribution based on the prefactors of the according final state vector in the computational basis.
This bit string is then used to calculate the simulated result of a projective measurement of a certain Pauli string.

We performed $M=1000$ measurement shots for every Pauli string that we evaluate and took the average of these to calculate the expectation value of the Hamiltonian for a specific system and geometry.
We are interested in statistical effects as well, hence, we repeated these steps $R=100$ times. Therefore, we find 100 different expectation values for all geometries and thus can determine the measurement variance for a measurement protocol relying on $M$ shots per operator.

\subsection{Influence of noise on N-representability}
\label{subsct:noise-influence}

Now we will discuss how the presented types of noise affect the validity of the five N-representability constrains listed in Sec.~\ref{sct:theory}.
First, we examine the quantum decoherence channels, i.e., dephasing, damping, and depolarization:

The effect of these channels will depend on how we encode the fermionic problem into qubits.
In this document, we rely on the Jordan-Wigner transformation, where we write:
\begin{subequations}
	\begin{align}
		c_k = Z_0 \otimes \cdots \otimes Z_{k-1} \otimes \sigma^+_k, \qquad \sigma_k^+ = \frac{1}{2}\left(X_k + \mathrm i Y_k\right), \label{eq:jw}\\
		c_k^\dagger = Z_0 \otimes \cdots \otimes Z_{k-1} \otimes \sigma^-_k, \qquad \sigma_k^- = \frac{1}{2}\left(X_k - \mathrm i Y_k\right).
		\label{eq:jw2}
	\end{align}
\end{subequations}
In these equations, $X_k$, $Y_k$, and $Z_k$ are the Pauli matrices of qubit $k$.

The first two listed N-representability conditions, hermiticity and antisymmetry, are not affected by any quantum channel.
Due to the properties of the fermionic operators, or their representation as qubit operators (e.g., in the Jordan-Wigner encoding), these conditions suffice for the elements of the RDMs, e.g., $^1D^i_j = \braket{c_i^\dagger c_j} = \mathrm{Tr} ( \rho c_i^\dagger c_j)$.
This is true as long as the density matrix, $\rho$, is a valid physical quantum state and not, for example, a random matrix or a non-Hermitian density matrix.
The application of decoherence still results in a physical state and therefore the measurement of the fermionic operators will always reveal these fundamental properties.
This also means that this is independent of the encoding, as long as the resulting fermionic operators are valid.
Likewise, the condition of positive semidefiniteness remains intact, since the regarded quantum noise channels are completely positive maps~\cite{nielsen_quantum_2010}.

The final two listed constraints, the trace and contraction relations, on the other hand both rely on the particle number $N$ staying constant.
In the Jordan-Wigner encoding the particle number $N$ is dependent on a sum of Pauli $Z$ matrices, since from Eqs.~\eqref{eq:jw} and~\eqref{eq:jw2} it follows that $c^\dagger_k c_k = \frac{1}{2} (1 - Z_k)$.
Dephasing noise, which -- as explained in above in Sec.~\ref{susct:decoherence} -- is essentially random $Z_k$ errors, therefore commutes with the particle number, i.e., $[Z_k, N] = 0$ for all $k$. Hence, using the Jordan-Wigner transformation, none of the N-representability constraints that we consider in this paper are violated by dephasing noise.
For this reason, in our numerical analysis that follows in Sec.~\ref{sct:results}, we do not study dephasing.
Note again, that this is due to our choice of encoding, using instead other fermion to qubit mappings like, e.g., the Bravyi-Kitaev transformation~\cite{bravyi_fermionic_2002}, the situation would be different, and dephasing would in fact violate this particular N-representability constraint.
In other words, our method is not fundamentally insensitive to dephasing noise.
Staying with the Jordan-Wigner encoding, the case is also different for damping and depolarizing noise, since, e.g., Pauli $X$ errors may occur, and $[X_k, N] \neq 0$, meaning the last two constraints can in fact be violated.
Consequently, these decoherence channels will be investigated in our numerics.

For shot noise the case is simple: It is completely statistical in its nature and may violate any of the five N-representability constrains mentioned above.

In the following, not all of the five constraints will be dealt with by post-processing.
Hermiticity and antisymmetry will be guaranteed by construction -- we will simulate measuring only the minimal necessary number of RDM elements and calculate the rest using the hermiticity and antisymmetry relations in Eq.~\eqref{eq:hermiticity}, \eqref{eq:hermiticity2}, and~\eqref{eq:antisymmetry}.
Positive semidefiniteness and the corrected trace will be attempted to enforce in a post-processing manner as explained in the subsequent Sec.~\ref{sct:method}.
Obtaining $^1D$ via the contraction in Eq.~\eqref{eq:contraction} is also nontrivial if the $^2D$ measurement is impaired by noise.
In order to perform the energy calculation according to Eq.~\eqref{eq:energy}, or perform the transformations to the two-hole or particle hole sector as in Eq.~\eqref{eq:trans1Q1D}, \eqref{eq:trans2Q2D}, and~\eqref{eq:trans2G2D}, without needing to first correct $^2D$, we simply simulate the measurement of $^1D$ additionally to the 2RDM; which is an insignificant overhead as the 2RDM contains quadratically more elements as the 1RDM.
Note that this approach gave us significantly better numerical results than trying to obtain $^1D$ solely via contraction.
In the future, one could try to improve this even further by taking into account the contractability constraint to jointly adjust $^1D$ and $^2D$ after measuring both.

\section{Procedure}
\label{sct:method}
In this section we explain how we calculated the data used to produce the plots and results presented in the following Sec.~\ref{sct:results}.
There, we have chosen to analyze three molecules as example systems, namely \ch{H2}, \ch{LiH} and \ch{BeH2}.
The first step was to derive the molecular Hamiltonian~\eqref{eq:ham}, where we represented \ch{H2} in the STO-3G basis, and \ch{LiH} as well as \ch{BeH2} in the MinAO basis~\cite{knizia_intrinsic_2013}.
Full configuration interaction (FCI) -- which is equivalent to exact diagonalization -- runs have been performed for all three systems, yielding our reference ground state $\ket{\psi_\mathrm{FCI}}$ and its respective energy $E_\mathrm{FCI}$ for each system.
Having obtained the state, using the Eqs.~\eqref{eq:1rdm} and~\eqref{eq:2rdm} given in Sec.~\ref{sct:theory} gives access to the respective RDMs $^1D_\mathrm{FCI}$ and $^2D_\mathrm{FCI}$.

Next we simulate decoherence, where we apply either damping, depolarizing, or dephasing noise to the reference state through the superoperator formalism described above in Sec.~\ref{sct:noise} (each with a rate $\Gamma = 10^{-2}$, relative to an evolution time $t = 1$).
Note again, that we use the Jordan-Wigner transformation to translate the fermionic operators to Pauli operators, which also defines the representation of the state and how the respective noise types affect it.
So we find by applying our superoperator to the initial density matrix $\rho_\mathrm{FCI} = \ket{\psi_\mathrm{FCI}}\bra{\psi_\mathrm{FCI}}$ the final density matrix $\rho_\mathrm{QC}$.
This (mixed) state $\rho_\mathrm{QC}$ mimics the result of a ground state calculation, for each of the example molecules, on a quantum computer, under the influence of decoherence.
With this state one can then derive the RDMs $^1D_\mathrm{QC}$ and $^2D_{QC}$, as well as consequently the respective energy $E_\mathrm{QC}$, where the index indicates that this is a simulated quantum computation.
For example, we do this by calculating ${^1D_\mathrm{QC}}^i_j = \braket{c_i^\dagger c_j}_\mathrm{QC} = \mathrm{Tr} (\rho_\mathrm{QC} c_i^\dagger c_j)$ if we don't simulate shot noise, where again we represent fermionic operators using the Jordan-Wigner transformation.
In the case of shot noise, the element is estimated by averaging over a number of samples of pure states (of the computational basis of the qubits). These samples are randomly chosen according to the probability distribution set by $\rho_\mathrm{QC}$ (which gives the same result as above in the limit of an infinite sample size).

Now we tried improving on the energy result by projecting the RDMs to the closest RDM that fulfills the N-representability constraints listed in Sec.~\ref{sct:theory}.
We guarantee that hermiticity and antisymmetry properties are valid by constructing the RDMs from measuring as little as necessary of the matrix elements and calculating the rest of the matrix via the respective relations.
Note that hermiticity and antisymmetry are not violated by applying decoherence channels, but this step will be important when we will consider shot noise below.

We have three options for projecting, fixing the two-particle RDMs, the two-hole RDMs, or the particle-hole RDM; these options we call D-projection, Q-pro\-jec\-tion, and G-projection, respectively.
To perform a D-projection, we take the measured particle 1-RDM ($^1D_\mathrm{QC}$) and the two-particle 2-RDM ($^2D_\mathrm{QC}$) and perform a fixed trace and positivity projection to both RDMs, which follows the algorithm from Ref.~\cite{rubin_application_2018} and is available as a function in the open source software \texttt{OpenFermion}~\cite{mcclean_openfermion_2020}. This function enforces the matrices to have a fixed trace -- which in this case is related to the particle number as given by Eqs.~\eqref{eq:trace1} and~\eqref{eq:trace2} -- and to be positive semi-definite.
For the Q-projection, we use $^1D_\mathrm{QC}$ and $^2D_\mathrm{QC}$ together with Eqs.~\eqref{eq:trans1Q1D} and~\eqref{eq:trans2Q2D} to obtain $^1Q_\mathrm{QC}$ and $^2Q_\mathrm{QC}$. These are then projected with the same function as in the case of the D-projection (fixing the number of holes instead of the number of particles). After the projection we transform back to the two-particle sector to evaluate the energy using Eq.~\eqref{eq:energy}.
Similarly, we proceeded for the G-projection, transforming from the two-particle to the particle-hole sector using Eq.~\eqref{eq:trans2G2D}, performing the projection, and then transforming back.

The same procedure is done again for simulations where we assume on top of decoherence also a finite amount of projective measurements, i.e., where the results are affected by shot noise (see Sec.~\ref{subsct:shot}).

Note again, that we measured $^1D_\mathrm{QC}$ additionally to $^2D_\mathrm{QC}$ (which is a negligible overhead).
Since the contraction from the 2-RDM to the 1-RDM following Eq.~\eqref{eq:contraction} is violated under the presence of noise, we use the measured $^1D$ elements to evaluate the energy $E_\mathrm{QC}$, and perform the transformation to the hole and particle-hole sectors according to Eq.~\eqref{eq:trans1Q1D}, \eqref{eq:trans2Q2D}, and~\eqref{eq:trans2G2D} to utilize the Q- and G-projections.
For the raw measurement, this yielded lower energy results than using a faulty contraction.
Furthermore, we observed that we could achieve lower energies for the individual Q- and G-projections; this is in comparison to using faulty transformations, and, in particular, also compared to otherwise needing to perform a D-projection first in order to allow for a reasonable contraction (which would be necessary for the transformations).

We evaluated in the case where we did not include shot noise the energy difference $\Delta E$ between the energy $E$, either of the raw measurement or after post-processing by on of the individual projections, versus the FCI energy for each geometry,
\begin{equation}
\label{eq:mae}
\Delta E = E - E_{\text{FCI}}.
\end{equation}
Furthermore, we investigated how close this procedure brings us to the FCI 2-RDM by looking at the fidelity:
\begin{equation}
\mathcal{F}({^2D}, {^2D}_{\text{FCI}}) = \left(\text{Tr}\sqrt{\sqrt{^2D}{^2D}_{\text{FCI}}\sqrt{^2D}}\right)^2,
\end{equation}
where $^2D$ describes the measured or once projected 2-RDM.
Note that the square roots are well defined as we are dealing only with positive semidefinite matrices in this case.

Including shot noise, we utilize the following paradigm:
We envision an experiment, where we assume resources to perform a total of $M = 1000$ measurement shots for every operator to measure; that is also for every data point in the plots in the following Sec.~\ref{sct:results}.
In order to analyze the statistics of performing such experiments, we repeat the same process $R = 100$ times.
For each data point, we therefore obtain $100$ different values $\mathcal{E}_i$ with $i \in \{1, ..., R\}$, where each $\mathcal{E}_i$ itself is the result of averaging over $1000$ shots (hence, for every point a total of $R \cdot M = 10^5$ shots are simulated).
Finally, we average over the repetitions, yielding
\begin{equation}
\mathcal E = \frac{1}{R}\sum_{i=1}^R \mathcal{E}_i,
\end{equation}
which is the energy we calculate again the energy difference to the reference energy:
\begin{equation}
\label{eq:maeShot}
\Delta E = \mathcal{E} - E_{\text{FCI}}.
\end{equation}

The reason for this averaging scheme is that, particularly for NISQ hardware, the number of measurements are a scarce resource.
Hence, we assume only $1000$ shots per operator in a single experiment.
Averaging again over multiple repetitions gives not only a more reliability expectation value to compare to the reference energy, but also allows to analyze the variance over the repetitions,
\begin{equation}
\label{eq:var}
\text{Var}(\mathcal{E})=\frac{1}{R}\sum_{i=1}^R\left(\mathcal{E}_i-\mathcal{E}\right)^2.
\end{equation}
This quantity gives insight about the expected accuracy of an energy measurement with only a limited number of $1000$ shots; specifically, it is interesting to which degree our proposed post-processing method lowers the variance.
Note, that there is a discussion on reducing the variance in Ref.~\cite{rubin_application_2018} using their method, which came at the expense of a small bias increase.
As we will see below, in our case this is not significant.
We will investigate a scenario, where shot noise will be an additional error source to decoherence, which we assume to be much stronger and as such to be the dominating systematic error.

In the following section we will show the results of our calculations.
There, we provide an analysis how individual projections improved on the above quantities for the different systems and noise types.
Furthermore, we will comment on approaches we tried to concatenate multiple projections to improve the results as much as possible, and how we propose to select the best projection method.

\section{Numerical analysis}
\label{sct:results}
After discussing the procedure above, here we show our data on how much improvement the D-, Q-, and G-projections grant on our simulated measurement results when dealing with shot and decoherence noise.
We examine if there is a preferred projection type for certain noise types, systems or geometries, and furthermore investigate how combining the different projection types alters the result.

The effects of damping and depolarization in addition to shot noise have been investigated on three systems, namely \ch{H_2}, \ch{LiH}, and \ch{BeH2}.
At first, simulations without shot noise will be discussed in Sec.~\ref{subsct:resNoShot} and afterwards the effect of shot noise will be included in Sec.~\ref{subsct:resShot}.
Besides the investigation of the energy deviations with respect to the FCI solution, we will furthermore take a look at the corresponding state fidelities for the individual projection methods.
The section discussing shot noise will provide figures with energy errors, as well as measurement variances.

\subsection{Simulations without shot noise}
\label{subsct:resNoShot}

\begin{figure*}
	\centering
	\subfloat[Energy errors for \ch{H2} under influence of damping noise. \label{fig:H2DampE}]{\includegraphics[width=5.9cm]{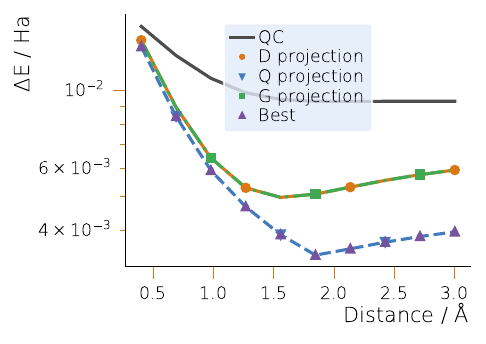}}\hfil
	\subfloat[Energy errors for \ch{LiH} under influence of damping noise. \label{fig:LiHDampE}]{\includegraphics[width=5.9cm]{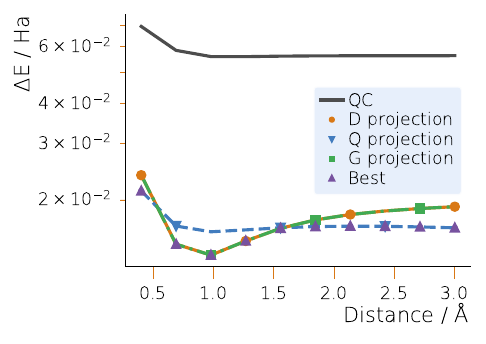}}\hfil 
	\subfloat[Energy errors for \ch{BeH2} under influence of damping noise. \label{fig:BeH2DampE}]{\includegraphics[width=5.9cm]{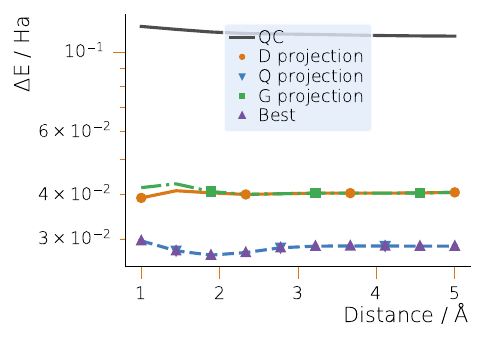}} 
	
	\subfloat[Fidelities for \ch{H2} under influence of damping noise. \label{fig:H2DampFid}]{\includegraphics[width=5.9cm]{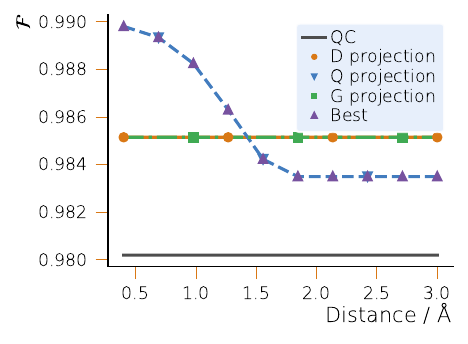}}\hfil   
	\subfloat[Fidelities for \ch{LiH} under influence of damping noise. \label{fig:LiHDampFid}]{\includegraphics[width=5.9cm]{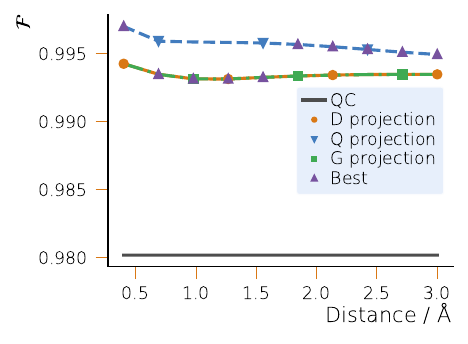}}\hfil
	\subfloat[Fidelities for \ch{BeH2} under influence of damping noise. \label{fig:BeH2DampFid}]{\includegraphics[width=5.9cm]{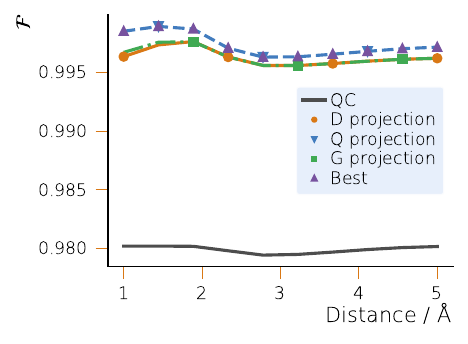}}
	
	\caption{Energy errors and fidelities for the three investigated examples \ch{H2}, \ch{LiH} and \ch{BeH2} for damping noise with a rate of $\Gamma=10^{-2}$. Shown are the results from measurements of the 1- and 2-RDM without post-processing, which are indicated as \textit{QC}, as well as the results from the single projections D, Q and G. \textit{Best} indicates the results yielding the best energies.
	}
	\label{Damping}
\end{figure*}

\begin{figure*}
	\centering
	\subfloat[Energy errors for \ch{H2} under influence of depolarizing noise. \label{fig:H2DepolE}]{\includegraphics[width=5.9cm]{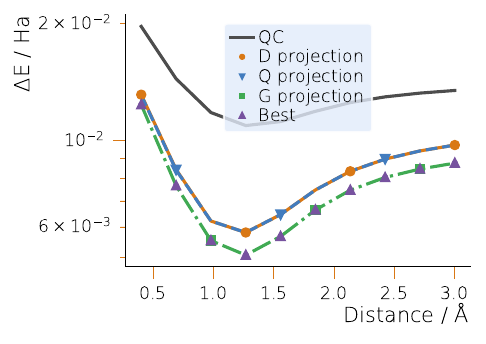}}\hfil
	\subfloat[Energy errors for \ch{LiH} under influence of depolarizing noise. \label{fig:LiHDepolE}]{\includegraphics[width=5.9cm]{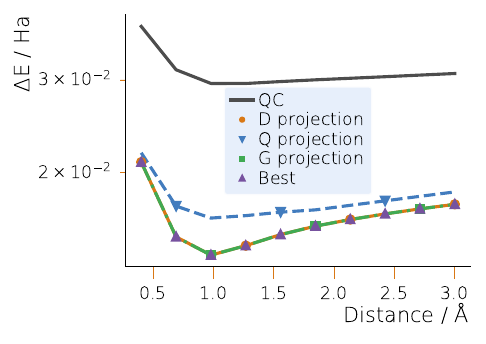}}\hfil 
	\subfloat[Energy errors for \ch{BeH2} under influence of depolarizing noise.  \label{fig:BeH2DepolE}]{\includegraphics[width=5.9cm]{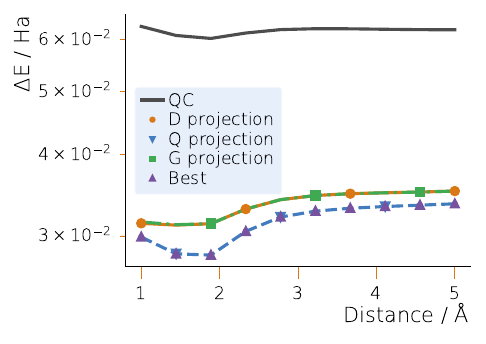}} 
	
	\subfloat[Fidelities for \ch{H2} under influence of depolarizing noise. \label{fig:H2DepolFid}]{\includegraphics[width=5.9cm]{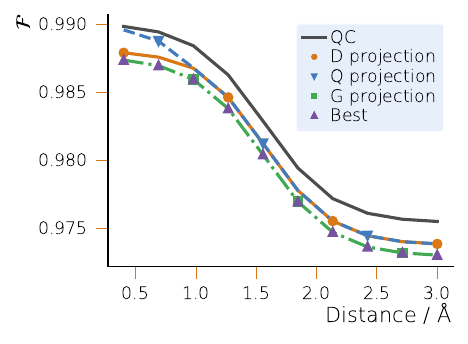}}\hfil   
	\subfloat[Fidelities for \ch{LiH} under influence of depolarizing noise. \label{fig:LiHDepolFid}]{\includegraphics[width=5.9cm]{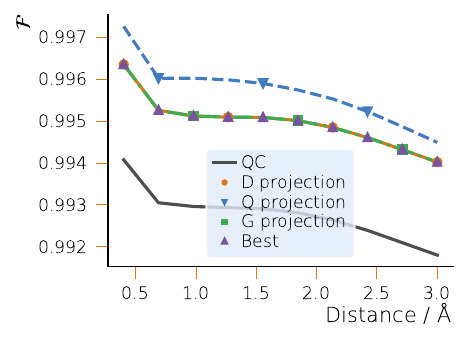}}\hfil
	\subfloat[Fidelities for \ch{BeH2} under influence of depolarizing noise. \label{fig:BeH2DepolFid}]{\includegraphics[width=5.9cm]{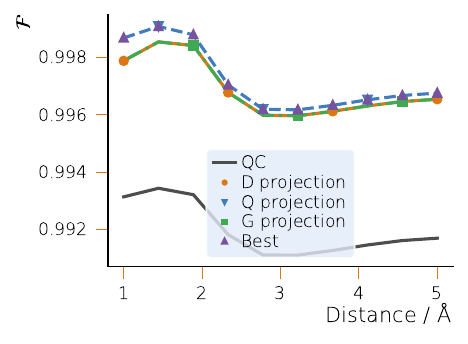}}
	
	\caption{Energy errors and fidelities for the three investigated examples \ch{H2}, \ch{LiH} and \ch{BeH2} for depolarizing noise with a rate of $\Gamma=10^{-2}$. Shown are the results from measurements of the 1- and 2-RDM without post-processing, which are indicated as \textit{QC}, as well as the results from the single projections D, Q and G. \textit{Best} indicates the results yielding the best energies.
	}
	\label{Depolarisation}
\end{figure*}

In this section we will investigate the energy errors and fidelities when performing simulations of the three systems with damping or depolarization present, but without shot noise, and compare the simulations without post-processing and with the three projection types to see how these improve the analyzed quantities.
Our simulation results are compiled in Fig.~\ref{Damping} and Fig.~\ref{Depolarisation}.

At first, we will analyze the data shown in Fig.~\ref{Damping}, where damping noise has been considered.
In Fig.~\ref{fig:H2DampE} to~\ref{fig:BeH2DampE} we plot the energy errors, and in Fig.~\ref{fig:H2DampFid} to~\ref{fig:BeH2DampFid} the fidelity, each time for \ch{H2}, \ch{LiH}, and \ch{BeH2}, respectively, at different inter-atomic distances.
For all cases, we plot the quantities derived from measurements of the 1- and 2-RDM without post-processing (labeled as QC), as well as after performing a single D-, Q-, or G-projection (labeled accordingly as D, Q, or G).
We also highlight the projection line that resulted in the smallest energy error (labeled as Best), making it easier to track the projection that yields the best energy.

From all sub-figures we can observe a significant improvement in terms of the energy error for all projection types; almost an order of magnitude lower errors are reached.
The data also verifies numerically that the projections do not fall below the FCI energy (which would be unphysical). We furthermore observe that certain projections lead to somewhat smaller energy errors. As for instance in Fig.~\ref{fig:H2DampE} it can be seen that the Q-projection leads to the best results for all distances between the hydrogen atoms, whereas the D- and G-projection lead to the same, worse result.
Interestingly, D- and G-projection lie on top of each other and the Q-projection deviates, which hints at a fundamental way damping affects the respective two-particle, particle-hole, and two-hole sectors.
However, it is dependent on the system, and even within a system dependent on the inter-atomic distance (see Fig.~\ref{fig:BeH2DampE}), which projection type yields the best energy.

Note, that these results qualitatively hold over a large range of decoherence rates $\Gamma$ (data is shown in Appx.~\ref{appx:noise-scaling}).
We acknowledge that this is the case even for very high noise rates of the order of the energy and prefactors in the Hamiltonian -- a regime where actual calculations to acquire numerically accurate results would not be sensible.
Going to even smaller noise rates than $\Gamma = 10^{-2}$, we see that we are already in a regime where the quantitative difference is basically a rescaling of the energy error based on the noise rate, i.e., the effect of decoherence is scaling linearly with the rate.
We should also note, that $\Gamma = 10^{-2}$ is already challenging to achieve for current NISQ devices; yet, in the results presented in this section, chemical accuracy is not quite reached.
For such desired accuracy, improvements in hardware would be needed, or one would need to combine the approach in this paper with further error mitigation methods.

The fidelity curves in Fig.~\ref{fig:H2DampFid} to~\ref{fig:BeH2DampFid} show general improvement of the fidelity performing post-processing, often times coming much closer to perfect fidelity than the initial QC result. We also observe the same behavior w.r.t.\ the D- and G-projection yielding the same value and Q deviating.
The absolute value of the fidelity as a rather abstract quantity is difficult to judge, but one way to interpret the data is to consider the difference of the fidelity from one, i.e., $1 - \mathcal{F}$.
We can think of this as some general error of the state, and we see that the D-, Q-, and G-projections reduce this error significantly, e.g., in the case of \ch{LiH} and \ch{BeH2} reducing it to roughly one fourth.
However, we see that the best projection in terms of the energy value does not necessarily yield the highest fidelity.
While from a heuristic argument one would expect lowering the energy error leads to approaching the correct 2-RDM as well, yet there is no direct connection of course.
This can be easily seen from considering examples with a dense lower spectrum, or even a degenerate ground state.
But it is important to realize that optimizing for the best energy will not guarantee all properties of the 2-RDM to be optimized as well.
If one is interested in quantities other than the energy, one might try to alternate the approach.

Comparing the above observations with the data in Fig.~\ref{Depolarisation}, where depolarizing noise was considered, we generally see similar results.
However, a few differences can be pointed out:
Which projection type yielded the best energy in the three systems (and for particular inter-atomic distances) is not equal to the case of damping noise.
We also do not observe anymore the strong link between the D- and G-projections (see Fig.~\ref{fig:H2DepolE}).
Hence, the best projection type does depend really on all variables considered; the chemical system, the inter-atomic distance, as well as the noise type, and there is no obvious a priori choice.
Another key difference to the previous figure is in Fig.~\ref{fig:H2DepolFid}, where we find that, unfortunately, the fidelity is actually reduced for all projections compared to the raw QC calculation.
This highlights on the other hand our advisory of being cautious when looking at properties other than the energy of the system.

Another point we would like to address is that so far we only looked at applying a single projection, either D, Q, or G.
In Refs.~\cite{lanssens_method_2018, rubin_application_2018} an iterative approach, where one applies one projection type after the other in an alternating fashion until the result converges, is proposed.
This is following the hope that in this way, the end result is as closely N-representable as possible using the projections at hand.

Pursuing this idea, we as well tried alternating sequences of the D-, Q-, and G-projections.
Following a projection with another one of a different kind in some cases changed (not necessarily lowered) the energy difference to the FCI reference.
However, in our systems we could not observe an improved energy when using a projection series versus the best energy result after only one projection.
Looking at the fidelity instead, we also could see quantitative changes, but not find a conclusive improvement in the sense that projection series would lead to higher fidelities.

On top of this analysis of projection series, we also tested if one can reach lower energy values if one performs \emph{partial} projections, possibly avoiding phenomena like local minima.
Here, we tried to iteratively post-process RDM's by only changing it towards the fully projected RDM by a small amount,
\begin{equation}
	^2D_{i+1} = \alpha B({^2D}_{i}) + (1-\alpha)({^2D}_{i}),
\end{equation}
where $B({^2D}_{i})$ is the two-particle RDM that stems from the energetically best projection (D, Q, or G) of ${^2D}_{i}$, $i$ is the iteration step, and $\alpha \in [0, 1]$ is the projection percentage.
However, we have observed that this procedure, even for very small values of $\alpha = 0.001$, converges again to the best result of simply one of the three projections.

Note that we found identical  results about the convergence of alternating projection series also for lower noise rates (see Appx.~\ref{appx:noise-scaling} how we are in a regime where lowering the noise basically linearly rescales its effect).
This suggests that the neighborhood around the perfectly $N$-representable RDMs that would be fix points for projections appears to be rather complex.
Our results are also not contradicting previous work, where the convergence criterion of the iterative procedure was based on the lowest eigenvalue of the RDM being close to zero (i.e., no negative eigenvalues).
Furthermore, the iterative results there seem to match fixed-trace projection method, which we rely on here as well.

We do not expect to undershoot the FCI energy when applying projections (note that the QC energy for decoherence will always be above the FCI energy, as decoherence leads to a physical state that is not the ground state).
However, we do not correct for all $N$-representability constraints in our projections, and we lack a formal proof that this cannot happen, but verified it in our extensive numerical analysis which is also consistent with previous work~\cite{rubin_application_2018}.%

Particularly since we are interested mostly in the energy calculation, and that in our data the projections do not yield unphysical energies below the FCI reference, we found the easiest approach for obtaining the best possible energy with the post-processing options available is to apply once the D-, Q, and G-projection and pick the best energy result.
This is a very simple to implement strategy that is furthermore very efficient computationally, particularly compared to iterative approaches.

\subsection{Simulations with shot noise}
\label{subsct:resShot}

\begin{figure*}
	\centering
	\subfloat[Energy errors for \ch{H2} under influence of damping and shot noise. \label{fig:H2DampShotE}]{\includegraphics[width=5.9cm]{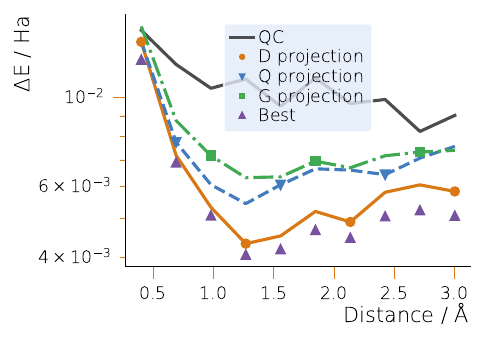}}\hfil
	\subfloat[Energy errors for \ch{LiH} under influence of damping and shot noise. \label{fig:LiHDampShotE}]{\includegraphics[width=5.9cm]{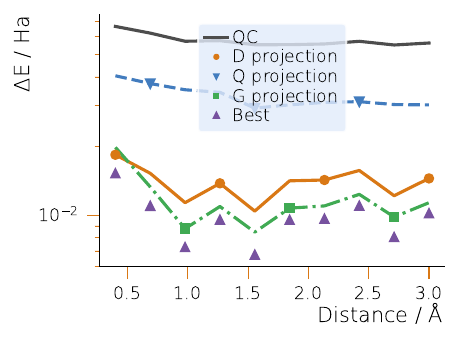}}\hfil 
	
	\subfloat[Variances for \ch{H2} under influence of damping and shot noise. \label{fig:H2DampShotVar}]{\includegraphics[width=5.9cm]{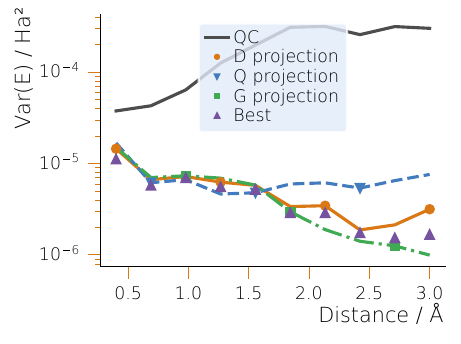}}\hfil   
	\subfloat[Variances for \ch{LiH} under influence of damping and shot noise. \label{fig:LiHDampShotVar}]{\includegraphics[width=5.9cm]{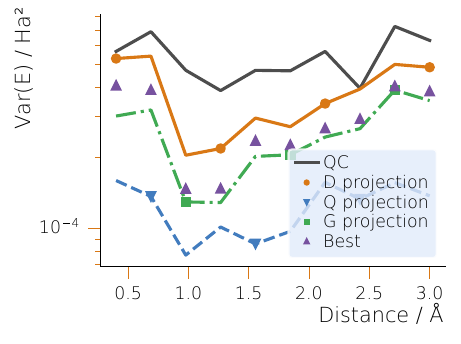}}\hfil
	
	\caption{Energy errors and measurement variances for the three investigated examples \ch{H2}, \ch{LiH} and \ch{BeH2} for damping and shot noise with 1000 measurement shots, 100 repetitions and a rate of $\Gamma = 10^{-2}$. Shown are the results from measurements of the 1- and 2-RDM without post-processing, which are indicated as \textit{QC}, as well as the results from the single projections D, Q and G. \textit{Best} indicates the results yielding the best energies.
	}
	\label{DampingShot}
\end{figure*}

\begin{figure*}
	\centering
	\subfloat[Energy errors for \ch{H2} under influence of depolarizing and shot noise. \label{fig:H2DepolShotE}]{\includegraphics[width=5.9cm]{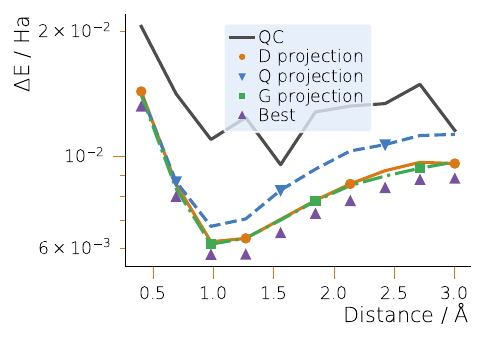}}\hfil
	\subfloat[Energy errors for \ch{LiH} under influence of depolarizing and shot noise. \label{fig:LiHDepolShotE}]{\includegraphics[width=5.9cm]{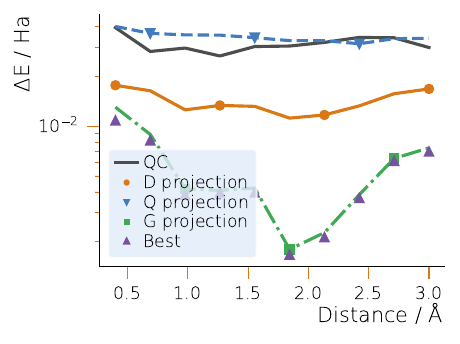}}\hfil 
	
	\subfloat[Variances for \ch{H2} under influence of depolarizing and shot noise. \label{fig:H2DepolShotVar}]{\includegraphics[width=5.9cm]{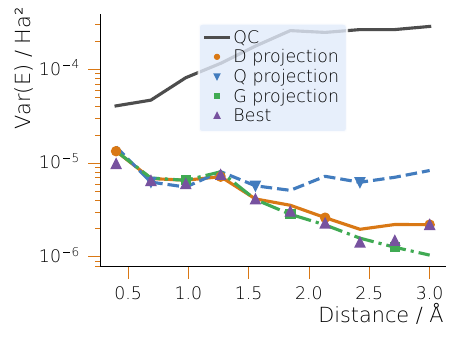}}\hfil   
	\subfloat[Variances for \ch{LiH} under influence of depolarizing and shot noise. \label{fig:LiHDepolShotVar}]{\includegraphics[width=5.9cm]{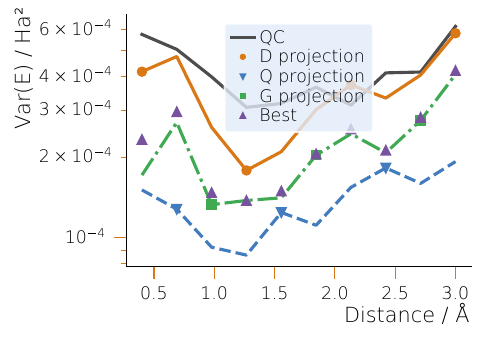}}\hfil
	
	\caption{Energy errors and measurement variances for the three investigated examples \ch{H2} and \ch{LiH} for depolarizing and shot noise with 1000 measurement shots, 100 repetitions and a rate of $\Gamma = 10^{-2}$. Shown are the results from measurements of the 1- and 2-RDM without post-processing, which are indicated as \textit{QC}, as well as the results from the single projections D, Q and G. \textit{Best} indicates the results yielding the best energies.
	}
	\label{DepolariseShot}
\end{figure*}

We now include effects of a finite number of measurements in the simulations of our chemical systems.
Though, we refrained from simulating the effect of decoherence in combination with shot noise for \ch{BeH2} due to time and resource constraints.
We simulated again damping in Fig.~\ref{DampingShot} and depolarization in Fig.~\ref{DepolariseShot}, now with the addition of shot noise stemming from measuring Pauli strings each with 1000 shots, furthermore averaging over 100 repetitions of such a scenario.
Note, that our investigated regime, the effect of decoherence was always strong compared to the statistical fluctuations from limited measurements.
This guaranteed in our simulations for energy results to remain above the FCI reference, which would not necessarily be the case for pure shot noise.
However, if one could drop below the FCI energy, our approach for picking the lowest energy projection as \emph{best} projection would not be meaningful.
On the other hand, in our chosen regime this is not an issue, and we believe that this regime of decoherence dominating shot noise is realistic to assume for NISQ applications.

Looking at Fig.~\ref{DampingShot} and Fig.~\ref{DepolariseShot}, we observe very similar results to Fig.~\ref{Damping} and Fig.~\ref{Depolarisation}:
We find a similar order of improvement in terms of energy error, and the best projection w.r.t.\ the energy is again not easily predictable but varies between systems, noise type and inter-atomic distance.
Note, that now the Best label does not follow a specific projection type; here, we choose the best projection w.r.t.\ the energy for every repetition individually.
Hence, for every repetition a different projection turns out to be the favored one, averaging over 100 repetitions yields a better value than for every other pure projection type.

Another difference to the plots of Sec.~\ref{subsct:resNoShot} is that instead of the fidelity we now plot the measurement variance as a relevant quantity in the context of measurement errors due to statistical shot noise.
Again, the points labeled Best do not necessarily present the best variance, as the optimization happened according to the energy.
But importantly, we see a significant reduction in the variance when post-processing the results by projecting to fulfill our selected N-representability constraints.
Specifically, in the case of larger distances in \ch{H2}, the reduction spans two orders of magnitude.

This variance reduction is a remarkable feature of the presented projection method.
It could potentially enable to measure quantities with a rather low number of shots while still remaining confident about the accuracy of the result.
Hence, the method could be a good candidate to reduce the measurement overhead which is a considerable obstacle in quantum computing, particularly in the NISQ era where quantum resources are fairly limited.
Studying this property of our or similar methods more extensively would be an interesting direction for future research.

\section{Conclusion}
\label{sct:conclusion}
The aim of this work was to try to reduce the energy error from 1- and 2-RDM calculations on a (simulated) quantum computer limited by decoherence and a finite number of measurement shots.
We investigated a post-processing method that enforced certain general N-representability constraints by projecting the measured RDMs into the subspace where these conditions were fulfilled.
Here, we regarded projecting the RDMs not just in the particle sector, but also the hole and particle-hole sector -- where one can switch between the sectors by simple transformations.

Specifically, we guaranteed hermiticity and antisymmetry by construction of the RDM from the measurements, and enforced positive semidefiniteness as well as the correct trace through the post-processing projection.
Analyzed were then ground state calculations of \ch{H2}, \ch{LiH} and \ch{BeH2} under the influence of damping and depolarizing channels, as well as shot noise.

We found the post-processing according to the N-representability constraints led to an improvement in terms of the energy error for all investigated example systems and noise types for all projections, i.e., the \mbox{D-,} Q- and G-projection in their respective particle, hole, and particle-hole sectors.
Similarly, the state fidelity was generally improved as well.
We could not observe an easily explained behavior which of the D-, Q-, or G-projection performed best depending on the system, the inter-atomic distance, or noise type.
On the other hand, we also found the best approach to always find the smallest energy error -- independent of the system or noise type -- is to simply take the smallest value of the three presented projection types.
Using approaches with series of alternating projections as previously suggested~\cite{lanssens_method_2018, rubin_application_2018}, or an iterative variant relying on partially projecting the RDM in each step, did not lead to better energy results; furthermore, they did not necessarily lead to the lower state fidelities either.
We note, that this simple way of finding the lowest energy hinges on the fact that we operate in a regime that is dominated by decoherence (versus shot noise), where we did not observe projecting to energies lower than the FCI reference.
However, this regime is reasonable to assume for NISQ devices.

In terms of the measurement variance, another investigated metric for shot noise simulations, we could see that the application of the proposed post-processing method led to a decrease by up to an order of magnitude.
The precise reduction in variance was again depending on the sector in which the projection was performed, without a clear choice to be made a priori.
While our approach of optimizing for the lowest energy did not necessarily yield the lowest measurement variance, it performed very well in general.
This makes it a viable approach for not just improving the energy error but also the variance; which is a particularly compelling feature, as this would enable to significantly increase the confidence in the accuracy of a quantum computation with a restricted number of measurement shots, especially considering the magnitude of the improvement.

In conclusion, we found our presented method of mitigating decoherence and shot noise to be very useful for improving energy calculations, particularly with respect to the highly effective reduction of the measurement variance.
In the considered noise regime, our practical approach to utilize the three different sectors to project proved not just to be simple but also fruitful.
The post-processing has low computational effort and there is no overhead to the quantum computation itself.
Our positive results spark interest for expanding on the method by including more constraints, e.g., further N-representability conditions, or other system-specific conserved symmetries.
Investigating these ideas we leave for future work.

\begin{acknowledgments}
This work was supported via the NEASQC pro\-ject funded by the European Union’s Horizon 2020 research and innovation program (grant agreement No.~951821).
We thank Nicolas Vogt for helpful discussions.

\end{acknowledgments}

\onecolumngrid

\section*{Appendices}

\begin{appendix}

\section{Superoperator example for two qubits}
\label{appx:multi-qubit}
	Assuming a general two-qubit density matrix $\rho_2$, we can vectorize it by stacking its columns into a single column vector,
	\begin{equation}
		\rho_2 = 
		\begin{pmatrix}
			\rho_{00,00} & \rho_{00,01} & \rho_{00,10} & \rho_{00,11} \\
			\rho_{01,00} & \rho_{01,01} & \rho_{01,10} & \rho_{01,11} \\
			\rho_{10,00} & \rho_{10,01} & \rho_{10,10} & \rho_{10,11} \\
			\rho_{11,00} & \rho_{11,01} & \rho_{11,10} & \rho_{11,11} 
		\end{pmatrix}
		\rightarrow
		\vec{\rho}_2 = 
		\begin{pmatrix}
			\rho_{00,00} \\
			\rho_{01,00} \\
			\rho_{10,00} \\
			\rho_{11,00} \\
			\vdots \\
			\rho_{11,11}
		\end{pmatrix},
	\end{equation}
	where the vectorization stacks the columns of the matrix in order. This results in a 16-dimensional column vector for two qubits.
	
	Next, we apply the two-qubit superoperator $\mathcal{L}_{2}$ to the vectorized density matrix:
	\begin{equation}
		\mathcal{L}_{2} \vec{\rho}_2 = \left( \mathcal{L}^{1} \otimes \mathcal{L}^{2} \right) \vec{\rho}_2,
	\end{equation}
	where $\mathcal{L}^1$ and $\mathcal{L}^2$ are single-qubit superoperators corresponding to the noise acting independently on each qubit. For example, if both qubits undergo the same dephasing rate, we use the corresponding superoperator
	\begin{equation}
		\mathcal{L} = 
		\begin{pmatrix}
			1 & 0 & 0 & 0 \\
			0 & \mathrm{e}^{-2\Gamma t} & 0 & 0 \\
			0 & 0 & \mathrm{e}^{-2\Gamma t} & 0 \\
			0 & 0 & 0 & 1
		\end{pmatrix},
	\end{equation}
	where $\Gamma$ is the dephasing rate and $t$ is the time the noise acts on the system.
	
	The total superoperator for the two-qubit case is then
	\begin{equation}
		\mathcal{L}_{2} = \mathcal{L} \otimes \mathcal{L},
	\end{equation}
	which is a $16 \times 16$ matrix acting on the vectorized density matrix of the two-qubit system. Transforming the result of $\mathcal{L}_{2} \vec{\rho}_2$ back to matrix form yields the final density matrix after the action of noise. Finally, we obtain a matrix that looks as follows:
	
	\begin{equation}
		\rho_2' =
		\begin{pmatrix}
			\rho_{00,00} & \mathrm{e}^{-2\Gamma t}\rho_{00,01} & \mathrm{e}^{-2\Gamma t}\rho_{00,10} & \mathrm{e}^{-4\Gamma t} \rho_{00,11} \\
			\mathrm{e}^{-2\Gamma t} \rho_{01,00} & \rho_{01,01} & \mathrm{e}^{-4\Gamma t} \rho_{01,10} & \mathrm{e}^{-2\Gamma t}\rho_{01,11} \\
			\mathrm{e}^{-2\Gamma t} \rho_{10,00} & \mathrm{e}^{-4\Gamma t} \rho_{10,01} & \rho_{10,10} & \mathrm{e}^{-2\Gamma t}\rho_{10,11} \\
			\mathrm{e}^{-4\Gamma t} \rho_{11,00} & \mathrm{e}^{-2\Gamma t}\rho_{11,01} & \mathrm{e}^{-2\Gamma t}\rho_{11,10} & \rho_{11,11}
		\end{pmatrix}
	\end{equation}
	
	The superoperator formalism scales naturally to $N$ qubits. In this case, the density matrix is of size $2^N \times 2^N$, and its vectorized form is a column vector of dimension $4^N$. The total noise superoperator becomes
	\begin{equation}
		\mathcal{L}_{N} = \bigotimes_{i=1}^{N} \mathcal{L}^i,
	\end{equation}
	where each $\mathcal{L}^i$ acts on the respective qubit's subspace. This allows to model the effect of independent noise on each qubit in large quantum systems.

\section{Results for different noise rates}
\label{appx:noise-scaling}
In the numerical calculation in the main text, we always set the noise rate to $\Gamma = 10^{-2}$.
Here, we show the effect of scaling the noise rate in Fig.~\ref{fig:noise-scaling}:

\begin{figure*}
    \centering
    \subfloat[Energy errors for \ch{H2} under influence of damping noise with $\Gamma = 10^{-4}$.]{\includegraphics[width=5.9cm]{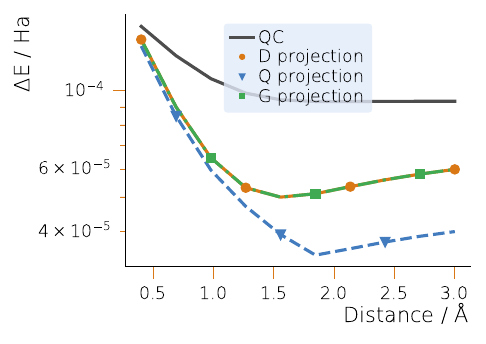}}
    \hfil
	\subfloat[Energy errors for \ch{H2} under influence of damping noise with $\Gamma = 10^{-2}$]{\includegraphics[width=5.9cm]{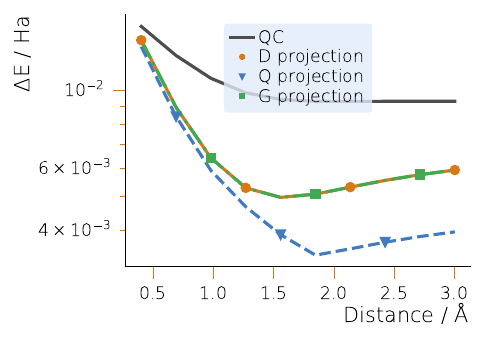}}\hfil 
	\subfloat[Energy errors for \ch{H2} under influence of damping noise with $\Gamma = 1$]{\includegraphics[width=5.9cm]{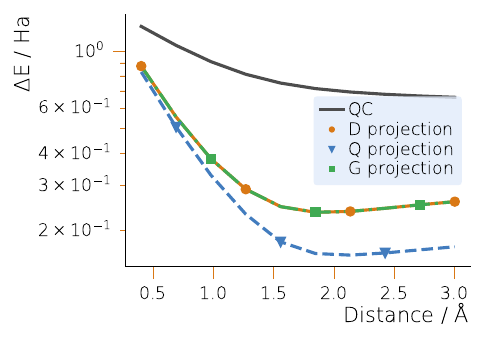}} 
    \caption{Energy errors for \ch{H2} under the influence of damping noise with different noise rates of: (a) $\Gamma = 10^{-4}$, (b) $\Gamma = 10^{-2}$, and (c) $\Gamma = 1$. Shown are the results from measurements of the 1- and 2-RDM without post-processing, which are indicated as \textit{QC}, as well as the results from the single projections D, Q and G.}
    \label{fig:noise-scaling}
\end{figure*}

As an example system, we focussed on \ch{H2} and considered damping noise.
Let us first compare the noise strength $10^{-2}$ from the main text with one reduced two orders of magnitude, i.e., $\Gamma = 10^{-4}$.
We realize, that we are already in a regime, where lowering the noise strength basically rescales the energy error proportionally, while the curve remains qualitatively the same.
Going to much higher noise rates of $\Gamma = 1$, we see the curve changing, but the main finding of the main text, that energy errors are lowered by roughly an order of magnitude, still holds.
We find similar results when scaling the noise rates of different noise types, and in different chemical systems.

Therefore, the results of the main text are robust within a large interval for the noise rate:
From really low decoherence (particularly for near future quantum devices), to a regime where the noise strength is comparable to the Hamiltonian evolution (where we would simulate noise as much as the actual system, i.e., a regime not sensible for numerically accurate quantum simulations of the chemical systems).

\end{appendix}

\twocolumngrid

\bibliography{nrep.bib,nrep2.bib}

\end{document}